%% file: br2.tex
\documentclass[12pt]{article}

\usepackage{amsfonts,amsmath}
\usepackage{latexsym}
\usepackage{times}
\usepackage{calc}
\usepackage{psfrag}
\usepackage[dvips]{graphicx}
\usepackage[T1]{fontenc} 
\usepackage[utf8]{inputenc} 
\usepackage[icelandic,english]{babel}
\usepackage{epsfig}
\usepackage{psfrag}
\usepackage{epsfig}
\input eqmacros
\input totalmacroe

\begin{document}

\selectlanguage{english}

 \topmargin 0pt
 \oddsidemargin 5mm
 \headheight 0pt
 \topskip 0mm

 \addtolength{\baselineskip}{0.5\baselineskip}

\pagestyle{empty}

\hfill

\vspace{1cm}

\begin{center}

{\Large \bf The spectral dimension of random brushes}

\medskip
\vspace{1.5 truecm} 


\vspace{1.5 truecm}

{\bf Thordur Jonsson and Sigurður Örn Stefánsson}

\vspace{0.4 truecm}


The Science Institute, University of Iceland

Dunhaga 3, 107 Reykjavik

Iceland

 \vspace{.7 truecm}

 \vspace{1.3 truecm}

\end{center}

\noindent {\bf Abstract.} We consider a class of random graphs,
called random brushes, which are constructed by adding linear graphs
of random lengths to the vertices of $\bbZ^d$ viewed as a graph. We
prove that for $d=2$ all random brushes have spectral dimension
$d_s=2$.  For $d=3$ we have ${5\over 2}\leq d_s\leq 3$ and for
$d\geq 4$ we have $3\leq d_s\leq d$.

 \newpage
 \pagestyle{plain}

\section{Introduction}
The generic structure of random geometrical objects is of interest
in many branches of physics ranging from condensed matter physics to
quantum gravity, see e.g.\ \cite{bookben} and \cite{book}.  One of
the methods used to analyze such objects is to study diffusion or
random walk. Diffusion allows us to define a notion of dimension,
the spectral dimension, for random geometrical objects. In recent
years the spectral dimension of triangulations has been studied
numerically in quantum gravity
\cite{earlynumsim,numerics,scaling,specdim1,specdim2} and
analytically for certain classes of random trees \cite{tjjw,DJW,GT}.
In \cite{DJW} the spectral dimension of various ensembles of random
combs was calculated. In this article we generalize the monotonicity
results of \cite{DJW} which allows us to find bounds on the spectral
dimensions of a class of graphs which we call brushes and define below.

Let $G$ be a connected, locally finite (i.e.\ each vertex has finitely many
nearest neighbours) rooted graph.  All graphs that we consider will be
assumed to have this property.  Let $p_G(t)$ be the probability
that a simple random walk on $G$ which starts at the root is back at
the root after $t$ steps.  If \beq{specdim0} p_G(t) \sim t^{-d_s/2}
\eeq as $t\to\infty$ then we say that $d_s$ is the spectral
dimension of the graph $G$. The existence of $d_s$ is not guaranteed
for individual graphs but its ensemble average can be shown to be
well defined in many cases \cite{DJW,GT}. It is easy to see 
that if the spectral
dimension exists then it is independent of the starting site of the
random walk.

Let us view $\bbZ^d$ as a graph with $j,k\in\bbZ^d$ neighbours if
their distance is 1 and let the origin of $\mathbb{Z}^{d}$ be the
root.  
It is well known that the spectral dimension of $\mathbb{Z}^{d}$ 
is $d$. Let $N_l$ be a linear chain of length $\ell$, i.e., the graph
obtained be connecting nearest neighbours in $\{0,1,{\ldots}
,\ell\}$ with a link.  Let $0$ be the root of $N_\ell$. Similarly,
let $N_\infty$ be the infinite linear chain with root at $0$. A
$d$-brush is a graph constructed by attaching one of the graphs
$N_\ell$ to each vertex of $\bbZ^d$ by identifying the root of
$N_\ell$ with a vertex in $\bbZ^d$, $\ell\in \bbN_0\cup \{\infty\}$,
$\ell =0$ corresponding to the empty chain. In a brush $B$ we will
refer to $\bbZ^d$ as the base and the linear chains as bristles.   A
random brush is defined by letting the length of the bristles be
identically and independently distributed by a probability measure
on $\bbN_0\cup \{\infty\}$. We see that the case $d=1$ corresponds
to the combs studied in \cite{DJW} which were shown to have a
spectral dimension in the interval $[1,{3\over 2}]$.

For $d>1$ 
we will show that the spectral dimensions of random brushes satisfy
the following: 
\bea \label{result} & d_s & =  2,~~~{\rm
if} ~~~d=2,\nonumber \\ {5\over 2}  \leq & d_s &  \leq 3,~~~{\rm if}
~~~d=3,\\  3 \leq & d_s &\leq  d,~~~{\rm if} ~~~d\geq 4.\nonumber
\eea 
Some  
comments are in order.  
We see that when $d\geq 3$, attaching the bristles to the base serves to lower
the spectral dimension since the spectral dimension of $\bbZ^d$ is
equal to $d$. This is opposite to the case of combs where the linear
chains tended to increase the spectral dimension. Intuitively this
can be understood in the following way. If there is a very long
bristle somewhere, a random walk can go up it and spend a long
time there before returning to the base which it must do eventually since
the bristles are recurrent.  Once it returns to the base
it will go back to the root with nonzero probability.  We will indeed
see below that adding a single infinite bristle to $\bbZ^d$ with
$d\geq 4$ will bring the spectral dimension down to $3$.
The two dimensional case is special because
$\bbZ^2$ is only marginally recurrent and the generating function 
for $p_{\bbZ^2}(t)$ has a logarithmic singularity which is not changed by the
presence of bristles.  Assuming that the spectral dimension of random
brushes can be calculated by mean field theory we show that the full
range of exponents in \rf{result} is realized.

The paper is organized as follows.  In the next section we define
the generating functions used to analyze the spectral dimension.
We then establish generalized monotonicity lemmas which are shown to
imply the stated bounds on $d_s$ in Section 4.  Section 5 contains a
discussion of mean field theory for brushes.  A final section contains 
some comments.

\section{Generating Functions}
Let $G$ be a graph and $p_G^1(t)$ 
the probability that a random walk is at the root at time $t$ for the
first time after $t = 0$. We define the return generating function
\begin {equation}
Q_G(z) = \sum_{t=0}^{\infty}p_G(t)z^t
\end {equation}
and the first return generating function
\begin {equation}
P_G(z) = \sum_{t=0}^{\infty}p^1_G(t)z^t.
\end {equation}
The generating functions are related by 
\begin {equation} \label{QPrelation}
Q_G(z) = \frac{1}{1-P_G(z)}.
\end {equation}
If $G$ has a spectral dimension $d_s$ then
\begin {equation} \label {singrelation}
Q_G^{(n)}(z) \sim \left\{ \begin{array}{ll}
 1 & \textrm{if $n = d_s/2-1$}\\
(1-z)^{d_s/2-1-n} & \textrm{otherwise}\\
\end{array} \right.
\end {equation}
where $n$ is the smallest nonnegative integer for which
$Q_G^{(n)}(z)$ diverges as $z \rightarrow 1$. Similarly, the behaviour (\ref{singrelation}) implies that the spectral
dimension is $d_s$. Here 
$f(y) \sim y^{\alpha}$ as $y \rightarrow 0$ means that for any $\epsilon > 0$ there exist
positive constants $c_1$ and $c_2$, which may depend on $\epsilon$,
such that
\begin {equation} \label {assym}
c_1y^{\alpha + \epsilon} \leq f(y) \leq c_2y^{\alpha - \epsilon}
\end {equation}
for $y$ small enough. Note that $f(y) \sim 1$ allows $f$ to have a logarithmic singularity at $0$.

The function $P_G(z)$ is analytic in the unit disc and $|P(z)| < 1$
for $|z| < 1$. If $P_G(z) \rightarrow 1$ as $z \rightarrow 1$ then $Q_G(z)$ clearly
diverges in which case the random walk is recurrent and $d_s \leq 2$. If $P_G(z)
\not\rightarrow 1$ as $z\rightarrow 1$ then the random walk is
transient and $d_s \geq 2$. In the latter case we see that if some derivative
$Q^{(n)}(z)$ diverges as $z\rightarrow 1$ then $Q_G^{(n)}(z) \sim
P_G^{(n)}(z)$ as $z \rightarrow 1$.

 If a graph has the property that every random walk which begins and ends at the root has an even number of steps, as is the case for brushes and bristles, we have to replace $p_G(t)$ with $p_G(2t)$ in (\ref{specdim0}) and $z$ with $z^2$ on the right hand side of (\ref{singrelation}). Then it is convenient to introduce a variable $x = 1 - z^2 \in [0,1]$. We will use the variable $z$ for general graphs but the variable $x$ when dealing with brushes and bristles.

We will need the following first return generating functions for the graphs $N_l$ and $N_\infty$ \cite{DJW}

\begin{equation}\label{linear}
P_l(x) = 1-\sqrt{x}\frac{(1+\sqrt{x})^l - (1-\sqrt{x})^l}{(1+\sqrt{x})^l + (1-\sqrt{x})^l} 
\end{equation}
and
\begin {equation} \label {geninfty}
P_\infty(x) = 1-\sqrt{x}.
\end {equation}

Let $\mu$ be a probability measure on
$\mathbb{N}_{0}\cup\{\infty\}$. Let $\mathcal{B}^d$ be the set of
all $d$-brushes.  We define a probability measure $\pi$ on
$\mathcal{B}^d$ by letting the measure of the set of $d$-brushes
$\Omega$ which have bristles at $n_1,n_2,...,n_k\in\bbZ^d$ of length
$\ell_1,\ell_2,...,\ell_k$ be
\begin {equation}
\pi(\Omega) = \prod_{i=1}^{k} \mu(l_i).
\end {equation}
The set $\mathcal{B}^d$ together with $\pi$ defines a random brush
ensemble. We
define the averaged generating functions
\begin {equation}
\overline{P}(x) = \langle P_B(x)\rangle_{\pi}
\end {equation}
and
\begin {equation}
\overline{Q}(x) = \langle Q_B(x)\rangle_{\pi}
\end {equation}
where $\langle \cdot \rangle_{\pi}$ denotes expectation with respect
to $\pi$. We say that a random brush has spectral dimension
$d_s$ if $\overline{Q}(x)$ obeys the relation (\ref{singrelation}) (after replacing $z$ with $z^2$ on the right hand side).

\section{Monotonicity} \label {chMon}

\newtheorem{theo}{Lemma}

Here we present the monotonicity results in a slightly more general setting
than is needed for the applications below.  This is both for clarity
and potential applications to random graphs different from
the brushes.

Let $G_1$ and $G_2$  be graphs such that $G_1$ can be
constructed from $G_2$ by attaching rooted graphs $F(i)$ by their
roots to sites $i\neq r$ of $G_2$. Let the roots of $G_1$ and $G_2$
be the same vertex (regarding $G_2$ as a subgraph of $G_1$).
The following result is a generalization of the Monotonicity Lemma of 
\cite{DJW}.

\begin {figure}[h]
\begin {center}
\includegraphics[width=0.6\textwidth]{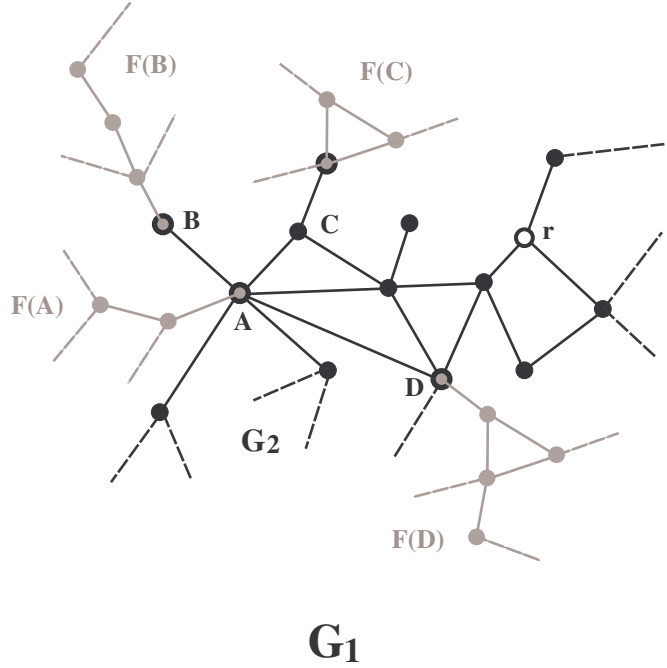}
\caption{An example of a graph $G_1$ constructed from $G_2$ and the
$F(i)$'s.}
\end {center}
\end {figure}

\begin {theo} \label{thmon1}

With $G_1$ and $G_2$ defined as above and $G_1 \neq G_2$ we have
\begin {equation} \label {mon1}
P_{G_1}(z) \leq P_{G_2}(z)
\end {equation}
with equality if and only if all the $F(i)$'s  are recurrent and z =
1.
\end {theo}
Proof: For any graph $G$ we can write $P_{G}(z)$ as a weighted sum over all random walks
$\omega$ on $G$ which start and end at the root without intermediate visits
to the root (this condition is denoted '$\omega$: FR on $G$'). Each walk $\omega$ has a weight 
\begin {equation}
W_G(\omega)=\prod_{t=0}^{|\omega|-1}\sigma_{G}(\omega_t)^{-1}
\end {equation}
where $\sigma_{G}(\omega_t)$ is the order of the vertex $\omega_t$
on $G$ where the walk is located at time $t$ and $|\omega |$ is
the number of steps in $\omega$. Each step of a walk has a factor $z$ associated with it so
\begin {equation} \label{f1}
P_{G}(z) = \sum_{\text{\scriptsize{$\omega$: FR on $G$}}}
W_G(\omega)z^{|\omega|}.
\end {equation}

Now consider a random walk $\omega'$ on $G_1$ which starts at the
root. Let $\omega$ be the subwalk of $\omega'$ which only travels on
$G_2$. If we look at the walk $\omega$ at time $t$ and location
$\omega_t$ then $\omega$ can be a subwalk of many different walks
$\omega'$ which correspond to all possible excursions into the graph
$F(\omega_t)$ before returning back to the walk on $G_2$. The weight
of these excursions is
\begin {equation} \label{factor1}
\sum_{n=0}^{\infty}\Big(\frac{\sigma_{F(\omega_t)}(\omega_t)}{\sigma_{G_1}(\omega_t)}P_{F(\omega_t)}(z)\Big)^n
=
\frac{1}{1-\Big(\frac{\sigma_{F(\omega_t)}(\omega_t)}{\sigma_{G_1}(\omega_t)}P_{F(\omega_t)}(z)\Big)}
\end {equation}
where $n$ counts the number of visits to $\omega_t$ before 
the walk leaves $\omega_t$ for another vertex on $G_2$ 
and the factor in front of
$P_{F(\omega_t)}(z)$ changes the order of the root of $F(\omega _t)$ to
$\sigma_{G_1}(\omega_t) = \sigma_{G_2}(\omega_t) +
\sigma_{F(\omega_t)}(\omega_t)$. The weight of the first step back
into $G_2$ after all the visits to $F(\omega_t)$ is
\begin {equation} \label {factor2}
\frac{z}{\sigma_{G_1}(\omega_t)}.
\end {equation}
Now replace the original weight
$\sigma_{G_2}(\omega_t)^{-1}z$ of $\omega$ at each point
$\omega_t\neq \omega_0$ by the product of the factors
(\ref{factor1}) and (\ref{factor2}). This newly weighted $\omega$
then accounts for every random walk on $G_1$ which has $\omega$ as a
subwalk on $G_2$. Thus we can write
\begin {eqnarray} \nonumber \label {rep}
P_{G_1}(z) &=& \sum_{\text{\scriptsize{$\omega$: FR on $G_2$}}}
\sigma_{G_2}(\omega_0)^{-1}z\prod_{t=1}^{|\omega|-1}\Big(\frac{z}{\sigma_{G_2}(\omega_t)+
\sigma_{F(\omega_t)}(\omega_t)(1
-P_{F(\omega_t)}(z))}\Big) \\
 &=& \sum_{\text{\scriptsize{$\omega$: FR on $G_2$}}}
K_{G_1,
G_2}(z;\omega)W_{G_2}(\omega)z^{|\omega|}
\end {eqnarray}
where in the last step we defined
\begin {equation}\label{KK}
K_{G_1,G_2}(z;\omega) =
\prod_{t=1}^{|\omega|-1}\Big(\frac{\sigma_{G_2}(\omega_t)}{\sigma_{G_2}(\omega_t)+
\sigma_{F(\omega_t)}(\omega_t)(1 - P_{F(\omega_t)}(z))}\Big).
\end {equation}
Since $P_{F(\omega_t)}(z) \leq 1$ with equality if and only if $F(\omega_t)$
is recurrent and $z=1$ it is clear that $K_{G_1,G_2}(z;\omega) \leq
1$ for all $z$ with equality if and only if all the graphs
$F(\omega_t)$ for a given $\omega$ on $G_2$ are recurrent and $z =
1$. The inequality
(\ref{mon1}) follows.
\begin{flushright} $\square$
\end{flushright}

\begin {theo} \label {thmon2}
Let $n\in\mathbb{Z}^+$ be such that $P^{(n-1)}_{G_2}(z)$
is continuous on the closed interval $[0,1]$. 
If all the $F(i)$'s are recurrent then
for a given $z\in]0,1[$ there exists a $\xi\in ]z,1[$ such that
\begin {equation}
P^{(n)}_{G_1}(\xi) \geq P^{(n)}_{G_2}(\xi).
\end {equation}
\end {theo}
Proof: We define
\begin {equation}
H_{G_1,G_2}(z;n) = \sum_{\text{\scriptsize{$\omega$: FR on $G_2$}}}
K_{G_1,G_2}(z;\omega)W_{G_2}(\omega)
\frac{d^{n-1}}{dz^{n-1}}z^{|\omega|}
\end {equation}
where $K_{G_1,G_2}$ is defined as above.  Every derivative of a (first) return generating function is a positive increasing function of $z\in[0,1[$ since the power series have no negative coefficients. It is easy to verify that the function $K_{G_1,G_2}$
has the same property. Therefore we get by
differentiating (\ref{rep}) $n$ times
\begin {eqnarray} \nonumber \label {long}
P^{(n)}_{G_1}(z) &=&
\sum_{i=0}^{n}\binom{n}{i}\sum_{\text{\scriptsize{$\omega$: FR on
$G_2$}}}K_{G_1,G_2}^{(i)}(z;\omega)W_{G_2}(\omega)
\Big(z^{|\omega|}\Big)^{(n-i)}
 \\ \nonumber &\geq& \sum_{\text{\scriptsize{$\omega$: FR on
$G_2$}}}K_{G_1,G_2}(z;\omega)W_{G_2}(\omega)
\Big(z^{|\omega|}\Big)^{(n)}
 \\ \nonumber &+& \text{\quad} n\sum_{\text{\scriptsize{$\omega$: FR on
$G_2$}}}K_{G_1,G_2}'(z;\omega)W_{G_2}(\omega)
\Big(z^{|\omega|}\Big)^{(n-1)}
\\ \nonumber &\geq& \sum_{\text{\scriptsize{$\omega$: FR on
$G_2$}}}K_{G_1,G_2}(z;\omega)W_{G_2}(\omega)
\Big(z^{|\omega|}\Big)^{(n)}
 \\ \nonumber&+& \sum_{\text{\scriptsize{$\omega$: FR on
$G_2$}}}K_{G_1,G_2}'(z;\omega)W_{G_2}(\omega)
\Big(z^{|\omega|}\Big)^{(n-1)}
\\ &=& H'_{G_1,G_2}(z;n).
\end {eqnarray}

With the same argument as in the proof of Lemma \ref{thmon1} it
holds that \\$H_{G_1,G_2}(z;n) \leq
P_{G_2}^{(n-1)}(z)$ with equality when $z=1$ since all
the $F(i)$'s are recurrent and because $P_{G_2}^{(n-1)}(z)$ and
therefore also $H_{G_1,G_2}(z;n)$ are continuous on $[0,1]$. Since $H_{G_1,G_2}(z;n)$ and
$P_{G_2}^{(n-1)}(z)$ are positive and increasing functions
of $z$ we find that
\begin {equation}
\frac{H_{G_1,G_2}(1;n)-H_{G_1,G_2}(z;n)}{P_{G_2}^{(n-1)}(1) -
P_{G_2}^{(n-1)}(z)} \geq 1.
\end {equation}
By a generalized mean-value theorem there exists a $\xi \in ]z,1[$
such that
\begin {eqnarray} \label {hospital}
\frac{H_{G_1,G_2}(1;n)-H_{G_1,G_2}(z;n)}{P_{G_2}^{(n-1)}(1) -
P_{G_2}^{(n-1)}(z)} = \frac{H'_{G_1,G_2}(\xi;n)}{P_{G_2}^{(n)}(\xi)}.
\end {eqnarray}
In view of (\ref{long}) the Lemma follows.
\begin{flushright} $\square$
\end{flushright}

\newtheorem {cor} {Theorem}

\begin {cor}  \label {corr}
Assume that all the $F(i)$'s are recurrent and that $G_1$ and $G_2$
have spectral dimensions $d_{s_1}$ and $d_{s_2}$ respectively. If
$G_2$ is recurrent then $G_1$ is recurrent and $d_{s_1} \geq
d_{s_2}$. If $G_2$ is transient then $G_1$ is transient and $d_{s_1}
\leq d_{s_2}$.
\end {cor}
Proof: Since all the $F(i)$'s are recurrent we have $P_{G_1}(1) = P_{G_2}(1)$ and therefore if $G_2$ is transient/recurrent then so is $G_1$.  First assume that $G_2$ is recurrent. Then by using Lemma 1 and Equations (\ref{QPrelation}), (\ref{singrelation}) and (\ref{assym})  we find that for any $\epsilon > 0$ there exist positive constants $c_1$ and $c_2$ which may depend on $\epsilon$ such that
\begin{equation}
c_1 (1-z)^{d_{s_1}/2-1+\epsilon} \leq Q_{G_1}(z) \leq Q_{G_2}(z) \leq c_2 (1-z)^{d_{s_2}/2 - 1 - \epsilon}
\end{equation}
for $z$ close to 1. If $d_{s_1} \neq d_{s_2}$ we choose $\epsilon < \frac{1}{4} |d_{s_2}-d_{s_1}|$ and send $z \rightarrow 1$ to conclude that $d_{s_1} > d_{s_2}$.  When $G_2$ is transient we use Lemma 2 and similar arguments as above to show that $d_{s_1} \leq d_{s_2}$.
\begin{flushright} $\square$
\end{flushright}

\section {The Spectral Dimension}
The $d$-brush where every bristle is $N_\infty$ we call 
the full $d$-brush and denote it $\ast d$.
We can relate the generating function of the full $d$-brush to the
generating functions of $\mathbb{Z}^{d}$ and $N_{\infty}$.  We use
the same argument as in the proof of Lemma 1.
Replacing all the graphs $F(i)$ with
$N_{\infty}$ and noting that the order of every point in
$\mathbb{Z}^{d}$ is $1/2d$  we get
\begin{equation} \label{f2}
P_{\ast d}(x) =
\Big(1+\frac{1-P_{\infty}(x)}{2d}\Big)P_{\mathbb{Z}^{d}}(x_{\text{ren}}(x))
\end {equation}
where $x_{\text{ren}}$ is defined by

\begin {equation} \label{f3}
\sqrt{1-x_{\text{ren}}} =
\frac{\sqrt{1-x}}{1+\frac{1-P_{\infty}(x)}{2d}}.
\end {equation}
We see that $x_{\text{ren}} = \sqrt{x}/d + O(x).$ By differentiating
(\ref{f2}) once and comparing with (\ref{singrelation}) we find the
spectral dimension of the full brush
\begin {equation} \label {BS}
d_{\ast} = \left\{ \begin{array}{ll}
 \frac{d}{2} + 1 & \quad\textrm{if $1 \leq d \leq 4$}\\
 3 & \quad\textrm{if $d \geq 4$.}\\
 \end{array} \right.
 \end {equation}
 If we replace the infinite bristles with finite ones, all of which
 have the same length, then with the same calculation we see that the
 spectral dimension remains equal to $d$. These are special cases of a
 more general result obtained in \cite{italir} for so called bundled
 structures. There, the base $\mathbb{Z}^{d}$ can be replaced by any
 graph $B$ and the infinite bristle (fiber) can also be replaced by
 any fixed graph $F$.
 
 Using the above calculation and Theorem 1 we can 
 find bounds on the spectral dimensions of fixed and random
 $d$-brushes.  Any fixed $d$-brush $B$ can be constructed from
 $\mathbb{Z}^{d}$ by attaching (recurrent) bristles to it and the
 full $d$-brush can be constructed from $B$ by attaching (recurrent)
 bristles to it. Therefore, by Theorem 1, 
 the spectral dimension of any fixed $d$-brush, if it
 exists, lies between $d$ and $d_{\ast}$.   This also holds for random
 brushes as is clear from equations (\ref{ramon1}) and
 (\ref{ramon2}) below and the proof of Theorem 1. 
 The spectral dimension for any fixed or random
 $d$-brush, if it exists, therefore obeys the inequalities
 (\ref{result}).
 
 The spectral dimension of random
 2-brushes always equals 2.  
 Indeed it follows from the fact that $Q_{\mathbb{Z}^{2}}(x)$ is asymptotic to
 $|\ln (x)|$ as $x\to 0$ and Lemma 1
 that there exist positive constants $c_1$ and $c_2$ such that
 \begin {equation}
 c_1|\ln(x)| \leq \overline{Q}(x) \leq c_2|\ln(x)|
 \end {equation}
 when $x$ is small enough . This is a stronger
 condition on the asymptotic behavior of $\overline{P}(x)$
 than $\overline{P}(x)\sim 1$ as
 $x\rightarrow 0 $.
 
 It is interesting that for $d \geq 4$ the lower bound on the
 spectral dimension always equals 3. In fact it is easy to see that
 attaching a single infinite bristle to $\mathbb{Z}^d$ with $d \geq
 4$ reduces the spectral dimension to 3. We can show this by
 attaching an infinite bristle to the root of $\mathbb{Z}^d$ since
 the spectral dimension is independent of the starting site of the
 random walks.  Let us call the resulting brush 
 $\bot d$.  The first return generating function is simply
 \begin {equation} \label{add}
 P_{\bot d}(x) = \frac{2d}{2d+1}P_{\mathbb{Z}^d}(x) +
 \frac{1}{2d+1}P_{\infty}(x).
 \end {equation}
  Since
  $d \geq 4$ equation (\ref{singrelation}) shows
  that $Q'_{\mathbb{Z}^{d}}(x)$ diverges slower than any negative
  power of $x$  as $x
  \rightarrow 0$ but
  $Q'_{\infty}(x) \sim x^{-1/2}$.   
  Therefore by differentiating (\ref{add}) we get
  \begin {equation} \label{Qprime}
  Q'_{\bot d}(x) \sim x^{-1/2}
  \end {equation}
 as $x \rightarrow 0$ and therefore by (\ref{singrelation}) the
 spectral dimension equals $3$.  It follows 
 that if a random $d$-brush with $d\geq 4$ has a nonzero probability
 of having one or more infinite bristles its spectral dimension
 equals 3.
 
 We find with similar arguments that adding a single (or finitely
 many) infinite bristles to $\mathbb{Z}^3$ gives the spectral dimension 3.
 However, if we add infinitely many bristles the spectral dimension
 of $\mathbb{Z}^3$ 
 can be lowered as is seen e.g.\ in the case of the full 3-brush.

We now use the notation of Section 3 and
consider the case when $G_2 = \mathbb{Z}^d$ and instead
of having a fixed $G_1$ we take a random $d$-brush. 
We would like to get bounds for the spectral dimension of
random brushes similar to those in Theorem 1. First
we note that by Lemma \ref{thmon1} we have for any
$B\in\mathcal{B}^d$ that
\begin {equation}
P_{\ast d}(x) \leq P_B(x) \leq P_{\mathbb{Z}^d}(x)
\end {equation}
and averaging we get
\begin {equation} \label {ramon1}
P_{\ast d}(x) \leq \overline{P}(x) \leq P_{\mathbb{Z}^d}(x).
\end {equation}
In order to generalize Lemma
\ref{thmon2} to random brushes we consider the case $d > 2$ and define the
functions
\begin {equation}
\overline{H}_a(x;n) = \langle H_{B,\mathbb{Z}^d}(x;n)\rangle_\pi \quad
\quad \text{and} \quad \quad \overline{H}_b(x) = \langle H_{\ast
d,B}(x;1)\rangle_\pi
\end {equation}
where $n=[{d-1\over 2}]$ is the smallest positive integer for which
$P_{\mathbb{Z}^d}^{(n)}(x)$ diverges as $x \rightarrow 0$. With the
same calculation as in (\ref{long}) we get
\begin {equation}
\frac{\overline{H}'_a(x)}{\overline{P}^{(n)}(x)} \leq 1 \quad \quad
\text{and} \quad \quad\frac{\overline{H}'_b(x)}{P'_{\ast d}(x)} \leq
1.
\end {equation}
We clearly have $(-1)^{n-1}\overline{H}_a(x) \leq
(-1)^{n-1}P_{\mathbb{Z}^d}^{(n-1)}(x)$ and $\overline{H}_b(x) \leq
\overline{P}(x)$ both with equality when $x=0$. Since the functions
$(-1)^{n-1}\overline{H}_a(x)$,$(-1)^{n-1}P_{\mathbb{Z}^d}^{(n-1)}(x)$,
$\overline{H}_b(x)$
 and
$\overline{P}(x)$ are all decreasing functions of $x$ we get with
the same argument as in the proof of Lemma 2 that for a given
$x\in]0,1[$ there exists a $\xi\in]0,x[$ such that
\begin {equation} \label {ramon2}
1 \leq \frac{\overline{P}^{(n)}(\xi)}{P_{\mathbb{Z}^d}^{(n)}(\xi)}
\quad \quad \text{and} \quad \quad 1 \leq \frac{P'_{\ast
d}(\xi)}{\overline{P}'(\xi)}.
\end {equation}
This extends Theorem 1 to random brushes and establishes the bounds
\rf{result}.

\section{Mean Field Theory}
It is an obvious question to ask whether the full range of spectral
dimensions allowed by \rf{result} is realized for some random brushes.
We do not have an answer to this question.  However, in \cite{DJW} the
spectral dimensions for different classes of random combs were
calculated exactly and shown to take the same values as in mean field
theory \cite{mft}.  By mean field theory we mean that the walk on the base (spine
in the case of combs) 
always sees a new bristle drawn from the
probability distribution $\mu$ whenever it is located at the root of a
bristle.  Since mean field theory is exact in one dimension we find it
likely that it is also exact in higher dimensions where the walks are
less likely to visit the same points on the base often.  Mean field theory
allows us to evaluate the spectral dimension very easily as we now
explain.

The ensemble average of the function $K_{G_1,G_2}$ defined in
\rf{KK} can be written
\begin {eqnarray} \nonumber
\langle K_{B,\mathbb{Z}^d}(x;\omega) \rangle_\pi &=&
\Big\langle \prod_{t=1}^{|\omega|-1}\frac{2d}{2d+
1 - P_{F(\omega_t)}(x)}\Big\rangle_\pi \\ &\stackrel{\text{\scriptsize{m.f.t.}}}{
=}& \Big(\Big\langle\frac{2d}{2d+
1 - P_{l}(x)}\Big\rangle_\mu\Big)^{|\omega| - 1}.
\end {eqnarray}
where the second equality is the mean field theory approximation.
The mean field theory approximation to the 
first return generating function is
\begin {equation} \label{mft}
 \overline{P}_{\text{\scriptsize{m.f.t.}},d}(x) =
\Big\langle\frac{2d}{2d+1-P_{l}(x)}\Big\rangle_\mu^{-1}
P_{\mathbb{Z}^{d}}(x_{\text{ren}}(x))
\end {equation}
where $x_{\text{ren}}(x)$ is defined through
\begin {equation}
\sqrt{1-x_{\text{ren}}(x)} =  \Big\langle\frac{2d}{2d+
1 - P_{l}(x)}\Big\rangle_\mu \sqrt{1-x}.
\end {equation}
Now choose $\mu(l) = c_a l^{-a}$ with $a>1$. The cases $d=1$ and $d=2$ we
understand.  Therefore 
consider the case $d\geq 3$.  It is straightforward to calculate 
the asymptotic behaviour of the following derivatives:
\begin {equation}
\langle P_l^{(n)}(x)\rangle_\mu  \sim x^{a/2-n} ~~~~~~~~~~~~ \text{for $n\geq 1$},
\end {equation}
\smallskip
\begin {equation} 
x_{\text{ren}}(x) \sim \left\{ \begin{array}{ll}
 x^{a/2} & \\
x & ,\\
\end{array} \right.\text{}~~~~~~~ ~~~~~~~ x_{\text{ren}}'(x) 
\sim \left\{ \begin{array}{ll}
 x^{a/2-1} & ~~~~~~\quad\textrm{if $1 < a \leq 2$}\\
1 & ~~~~~~\quad\textrm{if $a > 2$}\\
\end{array} \right.
\end {equation}
\smallskip
and 
\smallskip
\begin {equation} 
x_{\text{ren}}^{(n)}(x) \sim x^{a/2-n}  ~~~~~~~~~~~~ \text{for $n\geq 2$} 
\end {equation}
\smallskip
when $x \rightarrow 0$. We also see that the leading behaviour of the 
$n$-th derivative of (\ref{mft}) is 
\begin {equation}
\overline{P}^{(n)}_{\text{\scriptsize{m.f.t.}},d}(x) \sim \langle 
P_l^{(n)}(x)\rangle_\mu + P^{(n)}_{\mathbb{Z}^d}(x_{\text{ren}}(x))(x'_{
\text{ren}}(x))^n.
\end {equation}
First consider the case $d=3$, when we only have to look at the first 
derivative. Then $P'_{\mathbb{Z}^3}(x) \sim x^{-1/2}$ as 
$x\rightarrow0$ and therefore
\begin {equation}
\overline{P}'_{\text{\scriptsize{m.f.t.}},3}(x) \sim \left\{ \begin{array}{ll}
x^{a/4 - 1} & ~~~~~~\quad\textrm{if $1 < a \leq 2$}\\
x^{-1/2} & ~~~~~~\quad\textrm{if $a > 2$}\\
\end{array} \right.
\end {equation}
which gives
\begin {equation}
d_s = \left\{ \begin{array}{ll}
\frac{a}{2} + 2 & ~~~~~~\quad\textrm{if $1 < a \leq 2$}\\
3 & ~~~~~~\quad\textrm{if $a > 2$}.\\
\end{array} \right.
\end {equation}
Doing the same for $d\geq4$ we get the result
\begin {equation}
d_s = \left\{ \begin{array}{ll}
a + 2 & ~~~~~~\quad\textrm{if $1 < a \leq d-2$}\\
d & ~~~~~~\quad\textrm{if $a > d-2$}.\\
\end{array} \right.
\end {equation}
It is 
easy to see that putting a single bristle on $\mathbb{Z}^d$ with probability 
distribution $\mu$ for $d\geq 4$ gives the same spectral dimension as mean 
field theory.

Now consider the random brush defined by $\mu(\infty) = p > 0$ and $\mu(0) = 1-p$. It was shown in \cite{DJW} that for $d=1$ the spectral dimension of  this random brush equals the spectral dimension of the full brush. The same is of course true for $d=2$ and as well for $d\geq 4$, as was noted in the discussion below (\ref{Qprime}). Using mean field theory and similar analysis as above, we find that in any dimension the resulting random brush has also the same spectral dimension as the full brush. It is therefore clear that for this class of random brushes, if $d\neq3$, mean field theory gives the correct spectral dimension. Settling the case $d=3$ would require some extra work.

\section{Conclusions}
We have established bounds on the spectral dimensions of random graphs
constructed by attaching linear graphs to ${\mathbb Z}^d$ and argued
that mean field theory is likely to give the right value for the
spectral dimension.  The main monotonicity results are in fact valid
for a much larger class of graphs as explained in Section 3; the base
can be arbitrary and the bristles need only be recurrent graphs.  

While our random brushes do contain loops, they are all on the base
which is nonrandom and therefore do not yield much insight into how
one might hope to bound or evaluate the spectral dimension of random
graphs that contain loops like e.g.~random surfaces. For such graphs we need to develop new techniques.

\noindent {\bf Acknowledgment.}  This work is supported in part by
Marie Curie grant MRTN-CT-2004-005616, the Icelandic Science Fund
and the University of Iceland Research Fund.  We would like to thank
Bergfinnur Durhuus and John F.\ Wheater for many discussions and helpful comments on the manuscript.

\end{document}

%% file: eqmacros.tex
\def\void{}
\def\labelmark{}

\newenvironment{formula}[1]{\def\labelname{#1}
\ifx\void\labelname\def\junk{\begin{displaymath}}
\else\def\junk{\begin{equation}\label{\labelname}}\fi\junk}%
{\ifx\void\labelname\def\junk{\end{displaymath}}
\else\def\junk{\end{equation}}\fi\junk\labelmark\def\labelname{}}

{\ifx\void\labelname\def\junk{\end{array}\end{displaymath}}
\else\def\junk{\end{array}\right.\end{equation}}
\fi\junk\labelmark\def\labelname{}\def\junk{}
}

\newcommand{\beq}{\begin{formula}}
\newcommand{\eeq}{\end{formula}}
\newcommand{\beqv}{\begin{formula}{}}

%% file: totalmacroe.tex
\newcommand{\rf}[1]{(\ref{#1})}

\newcommand{\bea}{\begin{eqnarray}}
\newcommand{\eea}{\end{eqnarray}}
\newcommand{\beas}{\begin{eqnarray*}}
\newcommand{\eeas}{\end{eqnarray*}}
\newcommand{\beqs}{\begin{displaymath}}
\newcommand{\eeqs}{\end{displaymath}}

\newcommand{\ben}{\begin{equation}}
\newcommand{\een}{\end{equation}}

\newcommand{\bdm}{\begin{displaymath}}
\newcommand{\edm}{\end{displaymath}}

\newcommand{\bbN}{\mathbb N}
\newcommand{\bbZ}{{\mathbb Z}}